\documentclass{moriond}

\def\be{\begin{equation}}
\def\ee{\end{equation}}
\def\bea{\begin{eqnarray}}
\def\eea{\end{eqnarray}}

\newcommand{\Photo}{\includegraphics[height=35mm]{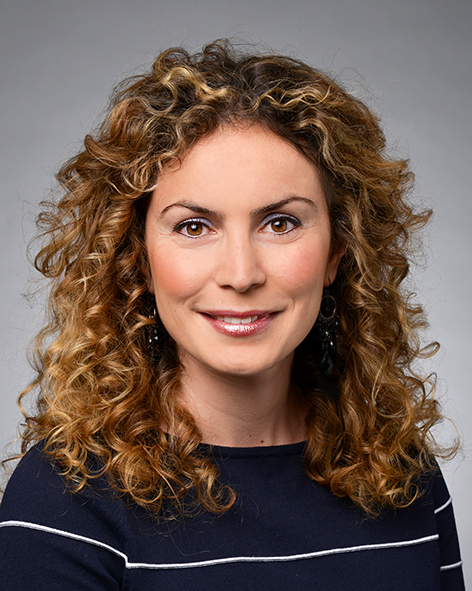}}

\begin{document}
\vspace*{4cm}
\title{GENERALISED PROCA THEORIES}

\author{ LAVINIA HEISENBERG }

\address{Institute for Theoretical Studies, ETH Zurich,\\
Clausiusstrasse 47, 8092 Zurich, Switzerland}

\maketitle\abstracts{
In this work we will summarise the recent progress made in constructing consistent theories for a massive vector field with derivative self-interactions. The construction is such that only the three desired polarisations of the Proca field propagate. We apply a systematic construction of the interactions by using the anti-symmetric Levi-Civita tensors. This finite family of allowed derivative self-interactions can be also obtained from the decoupling limit by imposing that the St\"uckelberg field only contains second derivatives both with itself and with the transverse modes. These interactions can be generalised to curved backgrounds, which relies on the presence of non-minimal couplings and constitutes a general family of vector-tensor interactions. We discuss also some extensions of these interactions by alleviating the restriction of second order nature of the equations or by imposing global symmetries. We will also comment on their interesting cosmological applications.}

\section{Introduction}

The Standard Model of Big Bang cosmology constitutes the prevailing cosmological model, which is able to satisfactorily represent the physics on cosmological scales using the two fundamental pillars of General Relativity and the Cosmological Principle. The latter stands for the homogeneity and isotropy. The combined inquiry of the available cosmological observations has firmly established $\Lambda$CDM as the standard model for our universe in a simple picture. It describes the evolution of the universe from its earliest known periods through the subsequent formation of large-scale structures and it provides comprehensive explanation for most of the observed phenomena. Despite being elegant and simple, the model relies on the  presence of three unknown ingredients, namely a cosmological constant acting as dark energy, a cold dark matter and an inflaton field, and it still faces some theoretical challenges. 
The most troublesome is the Cosmological Constant Problem and is a permanent reminder of our worrisome lack of a fully satisfactory and deep theoretical understanding for the value of the cosmological constant. Our theoretical foundation is shaky because we can not account for the enormous discrepancy between observations and the radiative corrections of massive particles to the vacuum energy using known standard techniques of quantum mechanics \cite{Weinberg:1988cp}. Along this line manifests itself another tenacious fundamental problem: we do not know how to construct a consistent theory of quantum gravity. Naive attempts of applying the principles of quantum mechanics to gravity immediately fail. The constructed theory of graviton bosons is not renormalisable and loses its predictive power. We are also prone to encounter classical primordial and black hole singularities. These singularities might be cured by the quantum nature of the interactions or by classical modifications of gravity at high curvatures (see for instance \cite{BeltranJimenez:2017doy}).

Concerning the dark matter component of the standard model, it is as essential and inevitable as the dark energy. It is fundamental for a successful description of the rotation curves of galaxies, Cosmic Microwave Background (CMB) anisotropies, large scale structures and weak lensing measurements. If dark matter exists, then we believe that it should be a distinctive matter from the ordinary baryonic matter beyond the standard model of particle physics. Its unique manifestation comes uniquely from its gravitational effect and it does not interact with photons. 
Notwithstanding the tremendous efforts, it has never been directly observed. Leaving aside the theoretical challenges, from purely observational side, the standard model with a cosmological constant and a cold dark matter fits the data almost impeccable. The word "almost" reflects the fact, that there are still some remaining anomalies. Just to mention a few, there is the tension in the Hubble constant obtained from local measurements versus from the CMB, the CMB hemispherical asymmetry together with the lack of power on large scales. There seem to be also some unusual large scale correlations and large scale bulk flows referred from distant quasar measurements. Even in the presence of dark matter, a slightly different cosmology is implied by the galaxy clusters than by the CMB measurements. It is also worth mentioning that there are unobserved predictions of the standard model, for instance the phase-space correlation of galaxy satellites and the generic formation of dark matter cusps in galaxies's central regions. On the contrary, the observations seem to indicate some tight correlations between the dynamical versus luminous mass, as the baryonic Tully-Fisher relation, which is not accounted for by the standard model. Most of the above mentioned anomalies do not have yet an overwhelming statistical significance, nonetheless, combined they might point out the failure of the assumptions of the Cosmological Principle and General Relativity. 

The above mentioned theoretical challenges together with the remaining observational anomalies have motivated the exploration of modifications of gravity both in UV and IR. Thanks to the advances in observational cosmology in high precision measurements, cosmology is the ideal place to test fundamental physics and get to the heart of the true nature of gravitational interactions. Most extensively studied modifications of gravity are based on scalar fields since they can for instance naturally give rise to accelerated expansion without changing the Cosmological Principle. A homogeneous and isotropic universe can be realised with a time dependent scalar field and the non-vanishing vev of the scalar field could naturally support a (quasi) de Sitter solution. However, as a candidate for dark energy the scalar field needs to be extremely light resulting unavoidably in long-range forces. Nonetheless, these additional fifth forces have never been observed in local gravity tests and therefore without any further mechanisms would be ruled out by observations. Luckily, screening mechanisms come to rescue. The presence of a successful screening mechanism makes it possible to disguise the scalar field on small scales while being unleashed on large scales producing the wanted cosmological effects. There are many modifications of General Relativity that bear an additional scalar field with these promising properties. Just to mention a few, for instance massive gravity \cite{deRham:2010ik,deRham:2010tw,deRham:2011by,Burrage:2011cr,deRham:2012ew} and higher dimensional frameworks \cite{Dvali:2000hr} naturally contain a scalar field as the helicity-0 part of the graviton with very specific non-linear interactions.

The Standard Model of elementary particles contains both abelian and non-abelian vector fields as the fundamental fields of the gauge interactions. Therefore, it is well motivated to explore the role of bosonic vector fields in the cosmological evolution. Traditionally, there is the worry that vector fields might generate large scale anisotropic expansion, which would make them not a natural candidate for dark energy. However, cosmic vector fields would naturally explain some of the aforementioned anomalies, specially those in relation with a possible preferred direction. A possible way to make the vector field to support isotropic cosmological solutions would be to promote the vector field to be a Proca field. A promising route in this respect has been considered in the works \cite{Heisenberg:2014rta,Allys:2015sht,Jimenez:2016isa} (see also \cite{Hull:2015uwa,Jimenez:2015fva,Khosravi:2014mua,Jimenez:2016opp,Jimenez:2013qsa}). These vector-tensor theories can indeed support isotropic cosmological solutions with the temporal component of the vector field \cite{Tasinato:2014eka,DeFelice:2016yws,Tasinato:2014mia,DeFelice:2016uil,Heisenberg:2016wtr,Lagos:2016wyv,deFelice:2017paw}. They also feature screening mechanisms \cite{DeFelice:2016cri}. Interesting extensions of these generalised Proca interactions can be constructed by alleviating some of the requirements or enriching some of the symmetries \cite{Heisenberg:2016eld,Kimura:2016rzw,Heisenberg:2016lux,Allys:2016kbq,Jimenez:2016upj,Hull:2014bga,Emami:2016ldl}. 

\subsection{Generalised Proca theories}\label{subsec:GeneralisedProcaTheories}
We would like to generalize the interactions of a massive vector field without changing the propagating number of degrees of freedom, namely two transverse and one longitudinal mode of the vector field. To start with, we can promote the mass term to a general potential term $V(A^2)$. This will not alter the spectrum of propagating degrees of freedom, since there is not any derivative of the vector field involved. In a similar way, we can consider any gauge invariant interactions constructed out of the field strength tensor $F_{\mu\nu}=\partial_\mu A_\nu-\partial_\nu A_\mu$ and its dual and also any contraction of those with the vector field, since they will not contain any dynamics for the zeroth component of the vector field $A_0$. Collecting all these type of interactions gives
\begin{equation}
\mathcal L_2  = f_2(A_\mu, F_{\mu\nu}, \tilde{F}_{\mu\nu}).
\end{equation}
The independent contractions will be in form of $X=-A_\mu A^\mu/2$, $F=-F_{\mu\nu}F^{\mu\nu}/4$ and $Y=A^\mu A^\nu F_{\mu}{}^{\alpha}F_{\nu\alpha}$ \cite{Heisenberg:2014rta,Fleury:2014qfa}and therefore this function can be also rewritten as $f_2(X,F,Y)$ (ignoring the partiy violating terms).
In first order in derivatives of $A_\mu$, we can start with the interaction
\begin{equation}
\mathcal L_3  = f_3(A^2)\;\; \partial\cdot A \label{vecGalL3} \,,
\end{equation}
with the arbitrary function $f_3(A^2)$. Note, that this interaction does not give a total divergence due to $f_3$. The temporal component of the vector field remains non-dynamical and the corresponding Hessian matrix vanishes identically. Alternatively, we can write this interaction in terms of the Levi-Civita tensor
 \begin{equation}
\mathcal{L}_3=-\frac{f_3(A^2)}{6}\epsilon^{\mu\nu\rho\sigma}\epsilon^{\alpha}{}_{\nu\rho\sigma}\partial_\mu A_\alpha=f_3(A^2)\partial\cdot A  \,.
 \end{equation}
 It is clear that there is only one way of contracting the indices of the Levi-Civita tensors at this order. This term would correspond to the cubic Galileon interaction for the longitudinal mode if we take the decoupling limit. The indices of the Levi-Civita tensors were contracted among themselves. We could have also contracted them with two additional vector fields, like so $\tilde{f}_3(A^2)\epsilon^{\mu\nu\rho\sigma}\epsilon^{\alpha\beta}{}_{\rho\sigma}\partial_\mu A_\alpha A_{\nu}A_{\beta}$. This would have resulted in an interaction that is conformally related to the previous one $\tilde{f}_3(A^2)A^\mu A^\nu (\partial_\mu A_\nu)$.
As next order in derivatives of the vector field we can consider the following three possible ways of contracting the Lorentz indices
\begin{equation}
\mathcal L_4  =  f_4(A^2) \; \left[c_1(\partial\cdot A)^2+c_2\partial_\rho A_\sigma \partial^\rho A^\sigma+c_3\partial_\rho A_\sigma \partial^\sigma A^\rho\right]\,,  \label{vecGalL4gen}
\end{equation}
where $f_4$ is again an arbitrary function and the parameters $c_1$, $c_2$ and $c_3$ need to be constrained in order to maintain the required property of three propagating degrees of freedom. In other words, we have to guarantee the presence of a second class constraint. We can ensure that by demanding that the corresponding determinant of the Hessian matrix $\det(H_{\mathcal L_4}^{\mu\nu})=2(c_1+c_2+c_3)(-2c_2)^3$ is zero. 
Hence, we need to ensure $c_1+c_2+c_3$ in order to obtain the required constraint. Choosing $c_1=1$ together with the condition $c_3=-(1+c_2)$, the Lagrangian then becomes
\begin{equation}
\mathcal L_4  =  f _4\; \left[(\partial\cdot A)^2+c_2\partial_\rho A_\sigma \partial^\rho A^\sigma-(1+c_2)\partial_\rho A_\sigma \partial^\sigma A^\rho\right]\,.  \label{vecGalL4}
\end{equation}
We can obtain these interactions in terms of the Levi-Civita tensors immediately. The anti-symmetric structure of the two tensors will directly impose the conditions that we just worked out by hand. Since the vector field has the symmetric and antisymmetric parts of $\partial_\mu A_\nu$ we can contract the indices in two ways
\begin{eqnarray}
\mathcal{L}_4&=&-\frac{1}{2}\epsilon^{\mu\nu\rho\sigma}\epsilon^{\alpha\beta}_{\;\;\;\;\rho\sigma}(f_4(A^2)\partial_\mu A_\alpha\partial_\nu A_\beta+c_2\tilde{f}_4(A^2)\partial_\mu A_\nu\partial_\alpha A_\beta)\nonumber\\
&=&f _4\; \left[(\partial\cdot A)^2-\partial_\rho A_\sigma \partial^\sigma A^\rho\right]+c_2\tilde{f} _4(\partial_\rho A_\sigma \partial^\rho A^\sigma-\partial_\rho A_\sigma \partial^\sigma A^\rho) \,.
\end{eqnarray}
One recognises that the terms proportional to $c_2$ are just the field strength tensor $F$
\begin{equation}\label{quartic_vecGal}
\mathcal L_4  = f _4 \; \left[(\partial\cdot A)^2-\partial_\rho A_\sigma \partial^\sigma A^\rho\right] + c_2 \tilde{f} _4F_{\rho\sigma}^2
\end{equation}
and can be thus absorbed into $f_2$. In a very similar way, we can construct the other interactions at next orders. We can either consider all the possible contractions of the interactions at each order and demand the vanishing of the determinant of the Hessian matrix, or we can directly apply the corresponding contractions with the Levi-Civita tensors. Both techniques result in the same terms. The self-interactions of the vector field can be summarised as the following Lagrangians at each order \cite{Heisenberg:2014rta,Jimenez:2016isa} 
\begin{eqnarray}\label{vecGalProcaField}
\mathcal L_2 & = &f_2(X,F,Y)\nonumber\\
\mathcal L_3 & = &f_3(A^2) \;\; \partial\cdot A \nonumber\\
\mathcal L_4  &=&  f _4(A^2)\;\left[(\partial\cdot A)^2-\partial_\rho A_\sigma \partial^\sigma A^\rho\right]   \nonumber\\
\mathcal L_5  &=&f_5(A^2)\;\left[(\partial\cdot A)^3-3(\partial\cdot A)\partial_\rho A_\sigma \partial^\sigma A^\rho 
+2\partial_\rho A_\sigma \partial^\gamma A^\rho\partial^\sigma A_\gamma \right]  +\tilde{f}_5(A^2)\tilde{F}^{\alpha\mu}\tilde{F}^\beta_{\;\;\mu}\partial_\alpha A_\beta \nonumber\\
\mathcal L_6  &=&f_6(A^2) \tilde{F}^{\alpha\beta}\tilde{F}^{\mu\nu}\partial_\alpha A_\mu \partial_\beta A_\nu \,.
\end{eqnarray}
Note, that the series stops after $\mathcal L_6$ and there are not any higher order interactions. All these interactions give rise to second order equations of motion for the vector field and by construction they do not give rise to any dynamics of the temporal component of the vector field \cite{Heisenberg:2014rta,Allys:2015sht,Jimenez:2016isa} .

\section{Alternative construction from the decoupling limit}
In a similar way we could have obtained the previously discussed systematic form of these interactions starting from the consistent interactions in the decoupling limit. In this limit the interactions of the transverse and longitudinal modes can be schematically written as an expansion
\be
\mathcal L\sim\sum_{m,n,p}c_{m,n,p}\left(\frac{A}{\Lambda_M}\right)^m\left(\frac{F}{\Lambda_F^2}\right)^n\left(\frac{S}{\Lambda_S^2}\right)^p \,,
\ee
where we introduced the symmetric part of the interactions as $S_{\mu\nu}=\partial_\mu A_\nu+\partial_\nu A_\mu$ and the scales $\Lambda_M$, $\Lambda_F$ and $\Lambda_S$ of the objects with some coefficients $c_{m,n,p}$. One very useful trick is to restore the broken gauge invariance of the vector field using the St\"uckelberg field as $A_\mu \to A_\mu+\partial_\mu\pi/M$, where $M$ stands for the mass of the vector field. In this language the scalar field $\pi$ represents the longitudinal mode of the original massive vector field. One can now very easily take the decoupling limit by sending $M\to0$ with the leading order contributions $A_\mu \to \partial_\mu\pi/M$ and similarly $S_{\mu\nu}\to\partial_\mu\partial_\nu\pi/M$. The aforementioned expansion of the interactions in this limit becomes
\be
\mathcal L_{\rm dec}\sim\sum_{m,n,p}c_{m,n,p}\left(\frac{\partial\pi}{M\Lambda_M}\right)^m\left(\frac{F}{\Lambda_F^2}\right)^n\left(\frac{\partial\partial\pi}{M\Lambda_S^2}\right)^p \,.
\label{Ldec}
\ee
We can now go order by order in the derivative of the scalar field and impose the conditions on $c_{m,n,p}$ after the summation. At the lowest order $p=0$, we only have one derivative per scalar field
\be\label{intDLp0}
\mathcal L_{\rm dec}^{p=0}\sim\sum_{m,n}c_{m,n,0}\left(\frac{\partial\pi}{M\Lambda_M}\right)^m\left(\frac{F}{\Lambda_F^2}\right)^n \,.
\ee
A glance at the interactions reveals that one would obtain three types of interactions at this order after summation in $m$ and $n$. The summation in $n$ maintaining $m=0$ would correspond just to functions of $F^2$. In the same way the summation in $m$ with $n=0$ would give rise to functions of the vector norm $A^2$. The third type of interactions at this order with $m\ne0$ and $n\ne0$ would just give rise to functions of $F_{\mu}{}^\alpha F_{\nu\alpha}A^\mu A^\nu$ type. Summarizing, the schematic interactions in (\ref{intDLp0}) correspond exactly to the interactions $f_2(X,F,Y)$ in (\ref{vecGalProcaField}) when one takes the decoupling limit. In the same way, we can analyse the next order interactions with $p=1$
\be\label{intDLp1}
\mathcal L_{\rm dec}^{p=1}\sim\sum_{m,n}c_{m,n,1}\left(\frac{\partial\pi}{M\Lambda_M}\right)^m\left(\frac{F}{\Lambda_F^2}\right)^n\left(\frac{\partial\partial\pi}{M\Lambda_S^2}\right) \,.
\ee
We will again have the interactions with $n=0$ and they represent nothing else but the standard Galileon interactions at cubic order $f_3(\partial\pi^2)\partial_\mu\partial^\mu\pi$. In fact, they represent the leading order interactions of $f_3(A^2) \;\; \partial\cdot A $ in $\mathcal L_3$. The novel interesting interactions arise for the case $n\ne0$, which gives rise to the first non-trivial mixing between the gauge field and the scalar field, where the scalar field comes in with second derivatives acting on it. After perfoming the summation in $m$ and $n$, the resulting symmetric rank-2 tensor will be contracted with $\partial_\mu \partial_\nu\pi$ and one has to impose that the $00-$ component of this tensor does not contain any time derivatives other than $\dot{\pi}$. 
In other words, this symmetric rank-2 tensor can only be built out of $f_3(\partial\pi^2)\tilde{F}^{\mu\alpha}\tilde{F}^\nu{}_\alpha$ in order to satisfy this requirement, since its magnetic part $\tilde{F}^{0\alpha}\tilde{F}^0{}_\alpha\propto B^2$ is purely potential. Hence, the decoupling limit interactions at this order can only be
\be
\mathcal L_{\rm dec}^{p=1}\sim\left(c_{2,0,1}(\partial\pi)^2\eta^{\mu\nu}+c_{0,2,1}\tilde{F}^{\mu\alpha}\tilde{F}^\nu{}_\alpha\right)\frac{\partial_\mu\partial_\nu\pi}{M\Lambda_S^2} \,.
\label{Ldecp1}
\ee
The coefficients can be arbitrary functions of $(\partial\pi^2)$. The first type of interactions are the leading order terms of $f_3(A^2) \partial\cdot A $ in $\mathcal L_3$ and the second type
interactions are the leading order contributions of $\tilde{f}_5(A^2)\tilde{F}^{\alpha\mu}\tilde{F}^\beta_{\;\;\mu}\partial_\alpha A_\beta$ in $\mathcal L_5$ in the decoupling limit. We see in the decoupling limit, that these are the only allowed interactions that give rise to second order equations of motion for both the scalar field and the gauge field. In complete analogy we can build the next order interactions with $p=2$
\be\label{intDLp2}
\mathcal L_{\rm dec}^{p=2}\sim\sum_{m,n}c_{m,n,2}\left(\frac{\partial\pi}{M\Lambda_M}\right)^m\left(\frac{F}{\Lambda_F^2}\right)^n\left(\frac{\partial\partial\pi}{M\Lambda_S^2}\right)^2 \,.
\ee
We have to impose again that the purely scalar interactions have to correspond to the scalar Galileon interactions whereas the mixed interactions with $n\ne0$ need to be constructed in a similar way as before, in other words the magnetic field of the gauge field should be allowed only to couple to the second time derivatives of $\pi$.
This leads uniquely to  
\begin{eqnarray}
\mathcal L_{\rm dec}^{p=2}\sim c_{2,0,2}(\partial\pi)^2\frac{(\partial_\alpha\partial^\alpha\pi)^2-(\partial_\mu\partial_\nu\pi)^2}{M^2\Lambda_S^4}
+c_{0,2,2}\tilde{F}^{\mu\nu}\tilde{F}^{\alpha\beta}\frac{\partial_\mu\partial_\alpha\pi\partial_\nu\partial_\beta\pi}{M^2\Lambda_S^4} \,,
\label{Ldecp2}
\end{eqnarray}
with the coefficients being arbitrary functions of  $(\partial\pi^2)$. The first type interactions is the purely quartic Galileon interactions, which are the leading order terms of the interactions $ f _4(A^2) \left[(\partial\cdot A)^2-\partial_\rho A_\sigma \partial^\sigma A^\rho\right] $ in $\mathcal L_4$ whereas the second type of interactions are the leading order mixed terms of $ f_6(A^2) \tilde{F}^{\alpha\beta}\tilde{F}^{\mu\nu}\partial_\alpha A_\mu \partial_\beta A_\nu$ in $\mathcal L_6$ in the decoupling limit. Finally, the cubic order interactions in $\partial\partial\pi$ with $p=3$ 
\be\label{intDLp3}
\mathcal L_{\rm dec}^{p=3}\sim\sum_{m,n}c_{m,n,3}\left(\frac{\partial\pi}{M\Lambda_M}\right)^m\left(\frac{F}{\Lambda_F^2}\right)^n\left(\frac{\partial\partial\pi}{M\Lambda_S^2}\right)^3 
\ee
contain only the pure Galileon interactions and there is no consistent mixing between the scalar field and the gauge field at this order. One can not construct any new interactions with $n\ne0$.
Note also, that the series stop here and hence the decoupling limit Lagrangian has a finite order of allowed interactions for the mixed couplings. From these decoupling limit Lagrangian one can obtain the full interactions by performing $\partial_\mu\pi\rightarrow A_\mu$ and $\partial_\mu\partial_\nu\pi\rightarrow S_{\mu\nu}$. They correspond to the same exact interactions that we constructed above using the Levi-Civita tensors. The total Lagrangian written in terms of the symmetric and antisymmetric parts is $\mathcal L_{\rm gen. Proca} = \sum^5_{n=2}\alpha_n \mathcal L^S_n$ with  
\begin{eqnarray}\label{vecGalProcaField}
\mathcal L^S_2 & = &f_2(A_\mu, F_{\mu\nu}, \tilde{F}_{\mu\nu})\nonumber\\
\mathcal L^S_3 & = &f_3(A^2)[S] \nonumber\\
\mathcal L^S_4  &=&  f _4(A^2)\;\left( [S]^2-[S^2] \right)   \nonumber\\
\mathcal L^S_5  &=&f_5(A^2)\;\left( [S]^3-3[S][S^2]+2[S^3]\right)  
+\tilde{f}_5(A^2)\tilde{F}^{\alpha\mu}\tilde{F}^\beta_{\;\;\mu}S_{\mu\nu} \nonumber\\
\mathcal L^S_6  &=&f_6(A^2) \tilde{F}^{\alpha\beta}\tilde{F}^{\mu\nu}S_{\alpha\mu}S_{\beta\nu} \,.
\end{eqnarray}
For the construction of these interactions the requirement of second order equations of motion and the absence of any dynamics for the temporal component of the vector field were very crucial. The latter condition reflects itself in the absence of higher order equation of motion for the scalar field once the broken gauge field is reintroduced using the St\"uckelberg trick. For more details see \cite{Heisenberg:2014rta,Jimenez:2016isa}. 
%%%%%%%%%%%%%%%%%
\section{Curved background}
The interactions we constructed above were so far on a flat background and it would be crucial to promote these interactions to the curved background case for different applications. In the presence of gravity the interactions have to be adjusted by counter-terms in order to maintain the property of second order equations of motion. If we simply replace the partial derivatives by covariant derivatives, then this would result in higher order equations of motion. In fact, specific non-minimal couplings have to be added for some of the interactions in order to guarantee the nature of second order equations of motion. For the construction of these non-minimal couplings the divergenceless tensors of the gravity sector play an important role. The pure St\"uckelberg field should this time possess the scalar Horndeski interactions \cite{Horndeski:1974wa}. This helps to construct the interactions for the vector field in curved backgrounds. The Lagrangian becomes $\mathcal L^{\rm curved}_{\rm gen. Proca}= \sqrt{-g}\sum^6_{n=2}\beta_n \mathcal L_n$ with \cite{Heisenberg:2014rta,Jimenez:2016isa} 
\begin{eqnarray}\label{vecGalcurv}
\mathcal L_2 & = & G_2(A_\mu,F_{\mu\nu}) \nonumber\\
\mathcal L_3 & = &G_3(X)\nabla_\mu A^\mu \nonumber\\
\mathcal L_4 & = & G_{4}(X)R+G_{4,X} \left[(\nabla_\mu A^\mu)^2-\nabla_\rho A_\sigma \nabla^\sigma A^\rho\right] \nonumber\\
\mathcal L_5 & = & G_5(X)G_{\mu\nu}\nabla^\mu A^\nu-\frac{1}{6}G_{5,X} \Big[
(\nabla\cdot A)^3 \nonumber\\
&+&2\nabla_\rho A_\sigma \nabla^\gamma A^\rho \nabla^\sigma A_\gamma -3(\nabla\cdot A)\nabla_\rho A_\sigma \nabla^\sigma A^\rho \Big] \nonumber \\
&-&g_5(X) \tilde{F}^{\alpha\mu}\tilde{F}^\beta_{\;\;\mu}\nabla_\alpha A_\beta  \nonumber \\
\mathcal L_6 & = & G_6(X)\mathcal{L}^{\mu\nu\alpha\beta}\nabla_\mu A_\nu \nabla_\alpha A_\beta 
+\frac{G_{6,X}}{2} \tilde{F}^{\alpha\beta}\tilde{F}^{\mu\nu}\nabla_\alpha A_\mu \nabla_\beta A_\nu \,,
\end{eqnarray}
where we have now the covariant derivatives $\nabla$ instead of partial and the vector field couples to the Ricci scalar, Einstein tensor and the double dual Riemann tensor $\mathcal{L}^{\mu\nu\alpha\beta}=\frac14\epsilon^{\mu\nu\rho\sigma}\epsilon^{\alpha\beta\gamma\delta}R_{\rho\sigma\gamma\delta}$ in order to counter balance the corresponding derivative interactions of the vector field. Note, that the interaction $\tilde{G}_5(Y) \tilde{F}^{\alpha\mu}\tilde{F}^\beta_{\;\;\mu}\nabla_\alpha A_\beta$ does not require the introduction of a non-minimal coupling.

\section{Extensions of Generalized Proca theories}
The interactions we constructed so far satisfy the condition of second order equations of motion for both the vector field and the graviton. One could alleviate this restriction and construct more general interactions. In accordance with the beyond Horndeski construction for scalar fields \cite{Zumalacarregui:2013pma,Gleyzes:2014dya}, one can construct similar beyond generalized Proca interactions allowing higher order nature of the equations of motion but still maintaining the correct number of propagating degrees of freedom \cite{Heisenberg:2016eld,Kimura:2016rzw}. The interactions need to be constructed such that the presence of a constraint equation is not jeopardised. With these less restrictive conditions one can for instance construct new beyond generalized Proca interactions ${\cal L}^{\rm N}={\cal L}_4^{\rm N}+{\cal L}_5^{\rm N}+
\tilde{{\cal L}_5^{\rm N}}+{\cal L}_6^{\rm N}$, where in terms of the Levi-Civita tensors their novelty becomes apparent \cite{Heisenberg:2016eld}
\begin{eqnarray}
\hspace{-1.2cm}
& &{\cal L}_4^{\rm N}
=f_4 (X)\hat{\delta}_{\alpha_1 \alpha_2 \alpha_3 \gamma_4}^{\beta_1 \beta_2\beta_3\gamma_4}
A^{\alpha_1}A_{\beta_1}
\nabla^{\alpha_2}A_{\beta_2} 
\nabla^{\alpha_3}A_{\beta_3}\,, \label{L4N}\\
\hspace{-1.2cm}
& &{\cal L}_5^{\rm N}
=
f_5 (X)\hat{\delta}_{\alpha_1 \alpha_2 \alpha_3 \alpha_4}^{\beta_1 \beta_2\beta_3\beta_4}
A^{\alpha_1}A_{\beta_1} \nabla^{\alpha_2} 
A_{\beta_2} \nabla^{\alpha_3} A_{\beta_3}
\nabla^{\alpha_4} A_{\beta_4}\,,\label{L5N} \\
\hspace{-1.2cm}
& &\tilde{{\cal L}}_5^{\rm N}
=
\tilde{f}_{5} (X)
\hat{\delta}_{\alpha_1 \alpha_2 \alpha_3 \alpha_4}^{\beta_1 \beta_2\beta_3\beta_4}
A^{\alpha_1}A_{\beta_1} \nabla^{\alpha_2} 
A^{\alpha_3} \nabla_{\beta_2} A_{\beta_3}
\nabla^{\alpha_4} A_{\beta_4}\,,
\label{L5Nd} \\
\hspace{-1.2cm}
& &
{\cal L}_6^{\rm N}
=f_{6}(X)
 \hat{\delta}_{\alpha_1 \alpha_2 \alpha_3 \alpha_4}^{\beta_1 \beta_2\beta_3\beta_4}
\nabla_{\beta_1} A_{\beta_2} \nabla^{\alpha_1}A^{\alpha_2}
\nabla_{\beta_3} A^{\alpha_3} \nabla_{\beta_4} A^{\alpha_4}\,,
\label{L6N}
\end{eqnarray}
where we introduced 
$\hat{\delta}_{\alpha_1 \alpha_2\gamma_3\gamma_4}^{\beta_1 \beta_2\gamma_3\gamma_4}=\epsilon_{\alpha_1 \alpha_2\gamma_3\gamma_4}
\epsilon^{\beta_1 \beta_2\gamma_3\gamma_4}$. The presence of these interactions results in a detuning between the relative coefficients of the non-minimal couplings to gravity
and the derivative self interactions of the vector field and gives rise to higher order equations of motion. However, due to the constraint equation one still has five propagating degrees of freedom, namely the two transverse graviton modes and the three vector modes.
Instead of using the systematical construction in terms of the Levi-Civita tensors, one can on a similar footing write down all the possible interactions between the fields and their derivatives with a priori arbitrary coefficients and put constraints on them afterwards coming from the vanishing of the determinant of the Hessian matrix as we did in \ref{subsec:GeneralisedProcaTheories} for the interactions in flat space-time. This reasoning was applied in \cite{Kimura:2016rzw} to construct the corresponding interactions up to $\mathcal{L}_4$ on curved space-time.

Another natural extension of the generalised Proca interactions can be constructed by promoting the broken gauge symmetry from $U(1)$ to the $SU(2)$ case. By breaking the non-abelian gauge symmetry, can we construct derivative self-interactions for a set of massive vector fields? One can indeed do that by following the same construction scheme using the antisymmetric Levi-Civita tensors and demanding that the equations of motion for $A_\mu^a$ remain second order while keeping the zero components non-dynamical \cite{Allys:2016kbq,Jimenez:2016upj,Hull:2014bga,Emami:2016ldl}. Some of these interactions will correspond to the straightforward generalisation of the single Proca field to the multi-Proca interactions with global $SO(3)$ symmetry \cite{Jimenez:2016upj}
\begin{eqnarray}\label{vecMGalcurv}
\mathcal L_2 & = & G_2(A_\mu^a,F^a_{\mu\nu}) \\
\mathcal L_4 & = & G_{4}R+G'_{4}\delta_{ab} \frac{ S^a_{\mu\nu}S^{b\mu\nu}-S^{a\mu}_\mu S^{b\nu}_\nu}{4}
\nonumber\\
\mathcal L_6 & = & G_6L^{\mu\nu\alpha\beta} F^a_{\mu\nu}F_{a\alpha\beta}
+\frac{G'_{6}}{2} \tilde{F}^{a\alpha\beta}\tilde{F}^{\mu\nu}_aS^b_{\alpha\mu} S_{b\beta\nu}. \nonumber
\end{eqnarray} 
It is worth to mention at this point that the restriction of the global $SO(3)$ symmetry diminishes the allowed interactions and for that reason one can not construct the analog of $\mathcal L_3$ and $\mathcal L_5$ interactions of the single Proca. Besides these direct extensions, there will be also genuine new interactions, which do not have the single Proca field analog \cite{Jimenez:2016upj}
\begin{eqnarray}
\mathcal L_3 & = & S^{a\mu\nu} A_\mu^b A_\nu^d A_\alpha^c A^{e \alpha} \delta_{de} \epsilon_{abc} \nonumber\\
\mathcal L_4&=&\epsilon^{\alpha\beta\gamma\delta}\tilde{F}^a_{\alpha\lambda} S^{b\lambda}{}_\beta A^a_\gamma A^b_\delta \nonumber\\
\mathcal L_5 & = &\epsilon_{abc} A^a_\mu A^{\mu d} \tilde{F}_d^{\alpha\nu}\tilde{F}^{b\beta}_\nu S^c_{\alpha\beta} \,.
\end{eqnarray} 
In difference to the interactions $\mathcal L_4 $ and $\mathcal L_6$ in (\ref{vecMGalcurv}), these genuine new interactions do not require the presence of non-minimal interactions. All these different extensions of the generalised Proca interactions offer promising routes to investigate further and may offer richer phenomenology. 

%%%%%%%%%%%%%%%%%%%%%%%%%%%%%%%%%
\section{Cosmological applications of generalized Proca theories}
If one deals with massless abelian gauge fields, then the only possible cosmological application can be realised by considering $N$ vector fields that are randomly distributed. In the case of a non-abelian vector field, one can construct interactions with global $SO(3)$ symmetry and consider field configurations where three vector fields point along the spatial directions. In the context of massive vector fields, one can achieve isotropic expansion with yet another field configuration. One can use the temporal component of the vector field as an auxiliary field in order to construct homogeneous and isotropic solutions. Compatible with the background of the metric $ds^2=-dt^2+a^2(t)d\vec{x}^2$, one can assume the following field configuration for the vector field:
\begin{equation}
A^{\mu}=(\phi(t),0,0,0)\,.
\end{equation}
From the action one can easily obtain the corresponding Einstein field equations for the background \cite{DeFelice:2016yws,DeFelice:2016uil}
\begin{eqnarray}
& &G_2-G_{2,X}\phi^2-3G_{3,X}H \phi^3
-6(2G_{4,X}+G_{4,XX}\phi^2)H^2\phi^2 \nonumber\\
& &+6G_4H^2+G_{5,XX} H^3\phi^5+ 5G_{5,X} H^3\phi^3
=\rho_M\,,
\label{be1}\\
& &
G_2-\dot{\phi}\phi^2G_{3,X}+2G_4\,(3H^2+2\dot{H})
-2G_{4,X}\phi \, ( 3H^2\phi +2H\dot{\phi}
+2\dot{H} \phi )\nonumber\\
&& -4G_{4,XX}H\dot{\phi}\phi^3+G_{5,XX}H^2\dot{\phi} \phi^4+G_{5,X}
H \phi^2(2\dot{H}\phi +2H^2\phi+3H\dot{\phi})
=-P_M\,,
\label{be2}
\end{eqnarray}
where $H=\dot{a}/a$ is the Hubble function. In a similar way, the vector field equations can be obtained in a straightforward way
\begin{equation}
\phi \left( G_{2,X}+3G_{3,X}H\phi +6G_{4,X}H^2
+6G_{4,XX}H^2\phi^2
-3G_{5,X}H^3\phi-G_{5,XX}H^3 \phi^3 \right)=0\,,
\label{be3}
\end{equation}
where $\rho_M$ and $P_M$ represent the energy density and pressure of the matter fields. One immediate observation is that the temporal component of the vector field
has only an algebraic equation and hence we can integrate it out in terms of the Hubble function. One also sees that the de Sitter solutions will correspond to $\dot{H}=0$
and $\dot{\phi}=0$, and will be attractor de Sitter fixed points. This type of cosmological solutions will be very relevant for the dark energy applications. However, because
of their attractor nature, these solutions will not be so interesting for the early universe applications. Instead, a better application will be delivered from the multi-Proca
generalisations. Apart from the standard field configurations $A^a_{i}=A(t)\delta^a_i$, as it was for instance considered in non-abelian gauge theories, one can assume
that the vector fields rather acquire the field configurations $A_\mu^a= \phi^a(t) \delta_\mu^0 + A(t) \delta_\mu^a$. These new field configurations open new promising possibilities 
for cosmological scenarios for the early universe, that have not been considered so far in the literature \cite{Jimenez:2016upj}.

\section*{Acknowledgments}

LH thanks financial support from Dr.~Max R\"ossler, the Walter Haefner Foundation and the ETH Zurich
Foundation.

\end{document}